\documentclass[a4paper,fleqn]{cas-dc}
\usepackage{graphicx}
\usepackage{printlen}
\usepackage{psfrag}
\usepackage{lipsum} 
\usepackage{amsmath}
\usepackage{float}
\usepackage{natbib}
\setcitestyle{round,comma,numbers,sort&compress,super}
\usepackage{tikz}

\usepackage{breakurl}

\renewcommand{\printorcid}{}

\usetikzlibrary{arrows,shapes,calc}
\usetikzlibrary{decorations.pathreplacing}

\newcommand\Rey{\mbox{\textit{Re}}}  
\newcommand\Pran{\mbox{\textit{Pr}}} 

\begin{document}

\let\WriteBookmarks\relax
\def\floatpagepagefraction{1}
\def\textpagefraction{.001}
\shorttitle{STREAmS-GPU}
\shortauthors{}

\title [mode = title]{STREAmS: a high-fidelity accelerated solver for direct numerical simulation of compressible turbulent flows}

\author{Matteo Bernardini$^1$, Davide Modesti$^2$, Francesco Salvadore$^3$, and Sergio Pirozzoli$^1$}

\address{$^1$Dipartimento di Ingegneria Meccanica e Aerospaziale, Sapienza Universit\`a di Roma, via Eudossiana 18, 00184 Roma, Italia}
\address{$^2$Aerodynamics Group, Faculty of Aerospace Engineering, Kluyverweg 2, 2629 HS Delft, The Netherlands}
\address{$^3$HPC Department, Cineca, Rome office, via dei Tizii 6/B, 00185 Roma, Italia}

\begin{abstract}
We present STREAmS, an in-house high-fidelity solver for large-scale, massively parallel direct numerical simulations (DNS)
of compressible turbulent flows on graphical processing units (GPUs).
STREAmS is written in the Fortran 90 language and it is tailored to carry out DNS of
canonical compressible wall-bounded flows, namely turbulent plane channel, zero-pressure gradient
turbulent boundary layer and supersonic oblique shock-wave/boundary layer interactions.
The solver incorporates state-of-the-art numerical algorithms, specifically designed to cope
with the challenging problems associated with the solution of high-speed turbulent flows
and can be used across a wide range of Mach numbers, extending from the low subsonic up to the
hypersonic regime.
The use of \verb+cuf+ automatic kernels allowed an easy and efficient porting on the GPU architecture
minimizing the changes to the original CPU code, which is also maintained.
We discuss a memory allocation strategy based on duplicated arrays for host and device which carefully minimizes the memory usage
making the solver suitable for large scale computations on the latest GPU cards.
Comparison between different CPUs and GPUs architectures strongly favor the latter, and executing the solver on a single NVIDIA Tesla P100
corresponds to using approximately 330 Intel Knights Landing CPU cores. 
STREAmS shows very good strong scalability and essentially ideal weak scalability up to 2048 GPUs, 
paving the way to simulations in the genuine high-Reynolds number regime, possibly at friction Reynolds number $Re_{\tau} > 10^4$. 
	The solver is released open source under GPLv3 license and is available at \url{https://github.com/matteobernardini/STREAmS}.
\end{abstract}


\maketitle
\section{Introduction}
Compressible flows are ubiquitous in aerospace applications 
and in recent years there has been a renewed interest in the field, 
owing to the rising investments in high-speed flight and space exploration.
These technological challenges call attention to 
high-fidelity numerical methods for compressible wall-bounded flows 
which have proved to be a valuable tool to unveil the complexity of these flows.

The flow physics of compressible wall-bounded turbulence is undoubtedly richer than that
of incompressible flows. The hyperbolic nature of the equations allows for the presence 
of propagating disturbances and discontinuities such as shock waves, which interact 
with the underlying turbulence, leading to flow phenomena that are absent in the incompressible case.
This additional complexity has affected and slowed down
the development of numerical methods for compressible flows, as compared to the incompressible ones.
Baseline numerical algorithms for direct numerical simulation (DNS) of incompressible flows were mainly developed
between the sixties and the eighties~\citep{patterson_71,kim_87,harlow_65,orlandi_00}, and basically settled since then.
The reliability of these algorithms and the advent of the open-source software promoted the development of several 
incompressible open-source solvers for fluid dynamics, both multi-purpose solvers as OpenFOAM~\citep{weller_98}, 
Nek5000~\citep{fisher_08} and Nektar++~\citep{cantwell_15} and 
academic solvers as AFiD~\citep{van_15} and CaNS~\citep{costa_18}.
These solvers are based on central processing units (CPUs) and message passing interface (MPI) parallelization,
which has been the standard approach in high-performance computing (HPC) \citep{lee_10} in the past twenty years. 
However, in the race towards exascale computing, the HPC architectures are showing consistent trend
towards the use of graphical processing units (GPUs).
In the last decade, GPUs have become the favorite solution to achieve accelerated cutting-edge performance
with high energy efficiency.
In particular, in the latest Top500 survey~\cite{dongarra_2011}, which reports the ranking
of the most powerful 500 machines worldwide,
136 machines are NVIDIA GPU-Accelerated \cite{nvidiablog_2019} for a total of about $40\%$ of the total power supplied.
In addition, NVIDIA GPUs power $90\%$ of the top 30 supercomputers on the Green500~\cite{green_500},
a list of HPC systems with high performance and improved energy efficiency. 
The incompressible DNS community  
has already benefited from improved computational performance of GPUs,
with two available in-house solvers AFiD-GPU~\citep{zhu_18} and CaNS-GPU~\citep{costa_20}.

Numerical algorithms for compressible flow DNS are less standardized then the incompressible ones,
as several formulations of the underlying equations are possible~\citep{honein_04,coppola_19}, 
each proving numerical advantages depending on the flow physics involved.
For this reason fewer CPUs-based open-source compressible flow solvers are available, compared to the incompressible case.  
Examples include popular multi-purpose open-source packages~\citep{weller_98,modesti_17,economon_15,cantwell_15} 
and OpenSBLI~\citep{jacobs_17}, a Python framework for the automated derivation of finite differences solvers
both for CPUs and GPUs architectures.
Another option is the use of the recent programming 
paradigm Legion~\citep{legion_2020} which allows to use the same solver
on different HPC architectures (including GPUs), without requiring extensive code restructuring.
A recent example of compressible flow solver using Legion is HTR~\cite{direnzo_20}, designed for hypersonic reacting flows.
To our best knowledge, no open-source GPUs-based compressible solver is currently available,
and the aim of this work is to fill this gap by adapting our CPUs-based compressible finite differences flow solver to run on multi-GPU clusters.
The CPUs solver stems from 20 years experience of our group on compressible wall-bounded flows and has been used
to carry out several seminal DNS studies of canonical flows including supersonic boundary
layer~\citep{pirozzoli_11,pirozzoli_13}, shock/boundary layer interaction (SBLI) \citep{pirozzoli_06,pirozzoli_10}, 
supersonic roughness-induced transition~\citep{bernardini_12} 
and supersonic internal flows~\citep{modesti_16,modesti_19,modesti_19duct}.
The solver was already ported to compute unified device architecture (CUDA) \citep{salvadore_13} for the
previous generation of GPUs, which required extensive code re-structuring and optimization.

In this work we present STREAmS (Supersonic TuRbulEnt Accelerated navier stokes Solver) 
a CUDA Fortran version of our compressible flow solver developed and optimized
for the latest generation of GPU clusters. 
We focus on three canonical wall-bounded turbulent flows, namely the supersonic plane channel,
the zero-pressure-gradient boundary layer developing over a flat plate and the oblique shock
wave/turbulent boundary layer interaction.
We discuss the CUDA implementation strategy and 
we compare the computational performance with the standard
CPU implementation and with the incompressible solvers AFiD-GPU and CaNS-GPU.

\section{Methodology}

STREAmS solves the fully compressible Navier-Stokes equations for a perfect heat-conducting gas
\begin{subequations}
 \begin{align}
  \frac{\partial \rho}{\partial t}   + 
  \frac{\partial\rho u_i}{\partial x_i}  & = 0 ,
  \label{eq:mass}\\
  \frac{\partial \rho u_i}{\partial t}  +
  \frac{\partial\rho u_i u_j}{\partial x_j} &= -
  \frac{\partial p}{\partial x_i}           +
  \frac{\partial\sigma_{ij}}{\partial x_j}  +
  f \delta_{i1} ,
  \label{eq:momentum}\\
  \frac{\partial \rho E}{\partial t} +
  \frac{\partial\rho u_j H}{\partial x_j} &= -
  \frac{\partial q_j}{\partial x_j}       +
  \frac{\partial\sigma_{ij}u_i}{\partial x_j}   +
  f u_1,
  \label{eq:energy}
 \end{align}
\end{subequations}
where $u_i$, $i=1,2,3$, is the velocity component in the i-th direction, $\rho$ the density,
$p$ the pressure, $E=c_v T+u_iu_i/2$ the total energy per unit mass, and $H=E+p/\rho$ is the total enthalpy.
The components of the heat flux vector $q_j$ and of the viscous stress tensor $\sigma_{ij}$ are
\begin{equation}
 \sigma_{ij}=\mu\left(\frac{\partial u_i}{\partial x_j} + \frac{\partial u_j}{\partial x_i} -\frac{2}{3} \frac{\partial u_k}{\partial x_k} \delta_{ij}\right), 
\end{equation}
\begin{equation}
 q_j=-k\frac{\partial T}{\partial x_j}, 
\label{eq:heat_vec}
\end{equation}
where the dependence of the viscosity coefficient on temperature is accounted for through Sutherland's law
and $k=c_p\mu/\Pran$ is the thermal conductivity, with $\Pran=0.72$.
The forcing term $f$ in equation \eqref{eq:momentum} is added in the plane channel flow simulations
and
is evaluated at each time step in order to discretely enforce constant mass-flow-rate in time. The
corresponding power spent is added to the right-hand-side of the total energy equation.

\subsection{Spatial discretization}

The convective terms in the Navier--Stokes equations are discretized using 
a hybrid energy-conservative shock-capturing scheme in locally conservative form~\citep{pirozzoli_10b}.
Let us consider the convective flux in one space direction (say $x$)
\begin{equation}
 f_x = \rho u \varphi,
\end{equation}
where $\varphi$ is the transported quantity, namely $\varphi = 1$ for the mass equation, 
$\varphi=u_j$ for the momentum equation and $H$ for the total energy equation.
The numerical discretization of the streamwise derivative of the flux $f_x$
on a uniform mesh with spacing $\Delta x$ relies on the identification of a numerical flux
$\hat{f}_{x\, i+1/2}$ defined at the intermediate nodes such that
\begin{equation}
\frac{\partial f_x}{\partial x}\bigg|_{i} =
\frac{1}{\Delta x} \left(\hat{f}_{x,\, i+1/2}-\hat{f}_{x,\, i-1/2}\right) .
\label{eq:num_flux}
\end{equation}
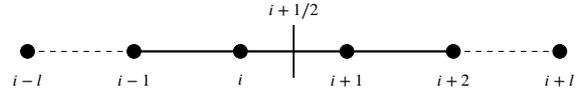
\begin{figure}
\begin{center}
\scalebox{0.7}{
\begin{tikzpicture}
 \coordinate(P1) at (0,0);
 \coordinate(P2) at (2,0);
 \coordinate(P0) at (-2,0);
 \coordinate(P3) at (4,0);
 \coordinate(P4) at (6,0);
 \coordinate(P5) at (-4,0);
 \coordinate (M) at ($(P1)!.5!(P2)$);
 \draw[fill=black] (P1) circle (0.4em) node [below=1em]{$i$};
 \draw[fill=black] (P2) circle (0.4em) node [below=1em]{$i+1$};
 \draw[fill=black] (P0) circle (0.4em) node [below=1em]{$i-1$};
 \draw[fill=black] (P3) circle (0.4em) node [below=1em]{$i+2$};
 \draw[fill=black] (P4) circle (0.4em) node [below=1em]{$i+l$};
 \draw[fill=black] (P5) circle (0.4em) node [below=1em]{$i-l$};
 \draw[line width=1.2pt] (P0)--(P1);
 \draw[line width=1.2pt] (P1)--(P2);
 \draw[line width=1.2pt] (P2)--(P3);
 \draw[dashed] (P3)--(P4);
 \draw[dashed] (P5)--(P0);
 \draw[line width=1.0pt] ($(M)+(0,0.5)$)--($(M)-(0,0.5)$);
 \draw ($(M)+(0,0.5)$)node [above]{$i+1/2$};
 \end{tikzpicture}
}
\caption{\label{fig:stencil} Sketch of the computational stencil in one space direction.}
\end{center}
\end{figure}

An energy-conserving numerical flux at the interface $i+1/2$ (figure~\ref{fig:stencil})
can be obtained by defining the three-point averaging operator~\citep{pirozzoli_10b}
\begin{equation}
 \left(\widetilde{F,G,H}\right)_{i,l} = \frac{1}{8}\left(F_i+F_{i+l}\right)\left(G_i+G_{i+l}\right)\left(H_i+H_{i+l}\right),
\label{eq:avg_operator}
\end{equation}
and recasting in conservative form the split formulation of the Eulerian fluxes~\citep{kennedy_08}
\begin{equation}
 \hat{f}_{x,\,i+1/2} = 2\,\sum_{l=1}^L a_l \sum_{m=0}^{l-1}\left(\widetilde{\rho,u,\varphi}\right)_{i-m,l} ,
\label{eq:encflux}
\end{equation}
where the $a_l$ are the standard coefficients for central finite-difference approximations of the first derivative,
yielding order of accuracy $2\,L$.
In smooth (shock-free) regions of the flow we use a fourth-order 
energy-consistent flux~\eqref{eq:encflux}, which guarantees that
the total kinetic energy is discretely  conserved in the limit case
of inviscid incompressible flow~\citep{pirozzoli_10}.
The locally conservative formulation allows
straightforward hybridization of the central flux with classical shock-capturing reconstructions.
In our case, shock-capturing capabilities rely on the use of Lax-Friedrichs flux vector splitting, whereby
the components of the positive and negative characteristic fluxes are reconstructed
at the interfaces using a weighted essentially non-oscillatory (WENO) reconstruction~\citep{jiang96}.
To judge on the local smoothness of the numerical solution we rely on a classical
shock sensor~\citep{ducros_99}
\begin{equation}
 \theta = \max{\left(\frac{-\nabla\cdot{u}}{\sqrt{{\nabla\cdot u}^2+{\nabla\times u}^2+u_{0}^2/L_0}},0\right)} \in[0,1],
\label{eq:sensor}
\end{equation}
where $u_0$ and $L_0$ are suitable velocity and length scales~\citep{pirozzoli_11}, defined
such that $\theta \approx 0$ in smooth zones, and $\theta \approx 1$ in the presence of shocks.
The viscous terms are expanded to Laplacian form
and also approximated with fourth-order formulas to avoid odd-even decoupling phenomena,
\begin{equation}
\begin{split}
\frac{\partial }{\partial x}\left(\mu\frac{\partial u}{\partial x}\bigg|_{i}\right)\bigg|_{i} = \frac{\partial \mu}{\partial x}\bigg|_{i}\frac{\partial u}{\partial x}\bigg|_{i} + \mu \frac{\partial^2 u}{\partial x^2}\bigg|_{i}=\\
\frac{1}{{\Delta x}^2}\sum_{l=-L}^L a_l^2 \mu_{i+l}u_{i+l} + \mu_i \frac{1}{{\Delta x}^2}\sum_{l=-L}^L b_l u_{i+l},
\end{split}
\end{equation}
where $b_l$ are the finite differences coefficient for the second derivative of order $2\,L$.

\subsection{Time integration}
A semi-discrete system of ordinary differential equations stems from discretization of
the spatial derivatives,
\begin{equation}
\frac{\mathrm{d}\mathbf{w}}{\mathrm{d}t} = \mathbf{R}(\mathbf{w})
\end{equation}
where $\mathbf{w}=[\rho, \rho u, \rho v, \rho w ,\rho E] $ is the vector of the conservative variables and $\mathbf{R}$ the vector of the residuals.
The system is advanced in time using Wray's three-stage third-order scheme~\citep{spalart_91},
\begin{equation}
 \mathbf{w}^{(\ell+1)} = \mathbf{w}^{(\ell)} +  \alpha_{\ell} \Delta t \mathbf{R}^{(\ell-1)} + \beta_{\ell} \Delta t \mathbf{R}^{(\ell)}, \quad \ell=0,1,2,
\label{eq:RK}
\end{equation}
$\mathbf{w}^{(0)}=\mathbf{w}^{n}$,
$\mathbf{w}^{n+1}=\mathbf{w}^{(3)}$
and the integration coefficient are $\alpha_{\ell} = (0, 17/60,-5/12)$, $\beta_{\ell} = (8/15, 5/12, 3/4)$.

\section{Validation}

STREAmS has been tailored to carry out three types of
canonical compressible flow configurations, 
namely supersonic plane channel flow,
supersonic boundary layer and
shock wave/boundary layer interaction.
In the following we validate the solver for these three flows
and compare the results to experimental and numerical data available 
in the literature. We use both Reynolds ($\phi=\overline{\phi} + \phi'$) and Favre ($\phi=\widetilde{\phi} + \phi''$,
$\widetilde{\phi}=\overline{\rho\phi}/\overline{\rho}$) decompositions,
where the overline symbol denotes averaging in the homogeneous space directions and in time.

\subsection{Supersonic plane channel flow}

\begin{figure*}
 \begin{center}
  \includegraphics[scale=0.75,clip]{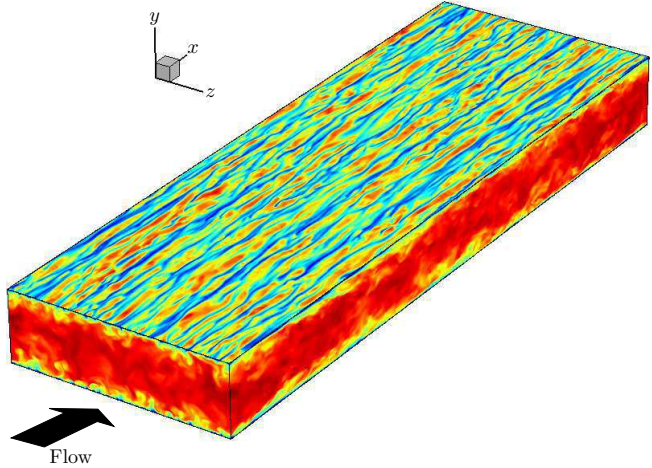}
  \caption{\label{fig:chann3d} Instantaneous streamwise velocity
  for plane supersonic channel flow at $\Rey_\tau=500$ and $M_b=1.5$. The wall-parallel plane is at $y^+=15$.}
 \end{center}
\end{figure*}

\begin{figure}
    \begin{center}
        \includegraphics[scale=0.95]{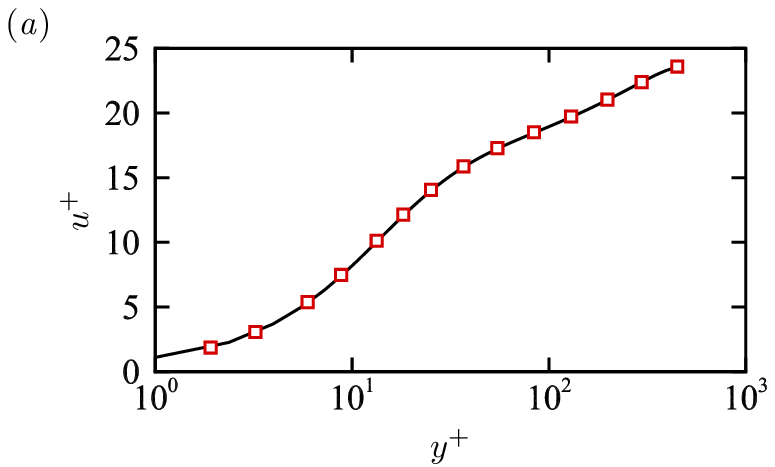}
        \includegraphics[scale=0.95]{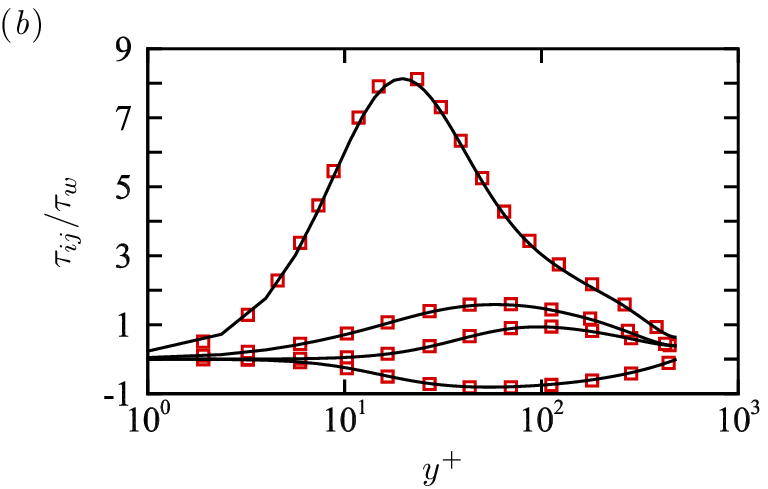}
	\caption{\label{fig:uchann} Supersonic plane channel flow at $M_b=1.5$
and $\Rey_\tau=490$. (\textit{a}) Mean streamwise velocity profile, 
     $u^+=u/u_\tau$ as a function of $y^+=y/\delta_v$. 
     (\textit{b}) Density scaled turbulent stresses $\tau_{ij}/\tau_w$ as a function of $y^+$.
     Present DNS data (black solid) are compared to previous DNS data obtained with the same numerical algorithm~\citep{modesti_16} (red dashed).}
    \end{center}
\end{figure}

We carry out DNS of plane supersonic channel flow at bulk Mach number $M_b=u_b/c_w=1.5$
and bulk Reynolds number $\Rey_b=2 \rho_b u_b h/\mu_w=14725$, where $\rho_b=1/V\int_V\rho\mathrm{d}V$ is the bulk density
and $u_b=1/(\rho_b V) \int_V \rho u \mathrm{d}V$
is the bulk velocity in the channel (both exactly constant in time), and $\mu_w$ and $c_w$ are the
dynamic viscosity coefficient
and the speed of sound at the wall temperature, respectively. 
This configuration corresponds to a friction Reynolds number $\Rey_\tau=\rho_w u_\tau h/\mu_w= 489$, where
$u_\tau=\sqrt{\tau_w/\rho_w}$ is the friction velocity and $\tau_w$ is the wall shear stress. 
The computational domain is a rectangular box with size $6 \pi h \times 2 h \times 2 \pi h$ in the $x,y,z$ coordinate directions, respectively and $h$ is the channel half-height.  
The mesh spacing is constant in the wall-parallel directions, and an error-function mapping is used
to cluster mesh points towards the walls. The number of mesh points in the three directions 
is $N_x=1024$, $N_y=256$, $N_z=512$, corresponding to a mesh spacing in wall units 
$\Delta x^+=9$, $\Delta y^+=0.8$--$5.7$ and $\Delta z^+=6$.
Periodicity is enforced in the homogeneous wall-parallel directions,
and no-slip isothermal conditions are imposed at the channel walls.
The mesh in the wall-normal direction is staggered such that 
the wall coincides with an intermediate node, where the convective fluxes are identically zero.
This approach guarantees correct telescoping of the numerical fluxes and 
exact conservation of the total mass,
with the further benefit of doubling the maximum allowed time step~\citep{modesti_16}.
The simulation is initiated with a parabolic streamwise velocity profile with superposed random perturbations
and large-scale sinusoidal perturbations, corresponding to streamwise-aligned rollers.
The channel flow simulation is carried out using the
central, energy-preserving flux only, as shock waves do not occur 
in this configuration.
Figure~\ref{fig:chann3d} shows the instantaneous streamwise velocity in a cross-stream, a streamwise 
and a wall-parallel plane at $y^+=y/\delta_v=15$. 
The instantaneous flow field exhibits the flow organization typical of incompressible wall turbulence, whereby
the wall-parallel plane is populated by low- and high-speed streaks associated with sweeps and ejections in the cross-stream plane.
Figure~\ref{fig:uchann} shows the mean velocity and Reynolds stress profiles.
Excellent agreement between result obtained with STREAmS-GPU and previous DNS carried out with 
CPU implementation of the solver~\citep{modesti_16} is found.


\subsection{Supersonic turbulent boundary layer}

We now consider a spatially-developing zero-pressure-gradient supersonic turbulent boundary layer evolving over a flat plate.
A direct numerical simulation is carried out at free-stream Mach number $M_{\infty} = 2$ and Reynolds number
in the low-moderate regime, up to a momentum thickness Reynolds number $Re_{\delta_2} \approx 1900$, corresponding to
a friction Reynolds number $Re_{\tau} \approx 600$.
As for the case of supersonic channel flow only the energy conservative flux is used as no shock wave discontinuities are present in the flow. 
To properly capture the large scale structures of the boundary layer (known as superstructures),
the simulation is carried out in a long and wide computational box,
which extends for $L_x = 105 \delta_{in}$, $L_y = 12 \delta_{in}$, $L_z = 10 \delta_{in}$,
in the streamwise (x), wall-normal (y) and spanwise (z) directions, $\delta_{in}$
being the boundary layer thickness at the inflow station, computed considering the $99\%$ of the free-stream velocity.
The computational domain is discretized with a mesh consisting of $N_x = 4096$, $N_y = 256$, $N_z = 512$ grid nodes.
Uniform mesh spacing is used in the wall-parallel directions, and hyperbolic sine stretching is used
in the wall-normal direction to cluster grid points towards to the wall, with wall spacing $\Delta y_w^+ = 0.8$.
The boundary conditions are specified as follows. At the upper and outflow boundaries non-reflecting
boundary conditions are imposed by performing characteristic decomposition in the direction normal to the boundary.
Similar characteristic wave treatment is also applied at the no-slip wall boundary, where temperature 
is set equal to its nominal recovery value
$T_r/T_{\infty} = 1+(\gamma-1)/2 \, r \, M_{\infty}^2$, with $r = Pr^{1/3}$.
In the spanwise direction the flow is assumed to be statistically homogeneous and periodic
boundary conditions are applied.
A critical issue in the simulation of spatially evolving turbulent flows is the prescription of the inflow turbulence
generation method. In STREAmS, velocity fluctuations at the inlet plane are imposed by means of a synthetic digital 
filtering (DF) approach~\citep{klein_03},
extended to the compressible case thanks to the use of the strong Reynolds analogy~\citep{touber09}.
An efficient implementation of the method is achieved using an optimized DF procedure~\citep{kempf12}, 
whereby the filtering operation is decomposed in a sequence of fast one-dimensional convolutions.
The implementation requires the specification of the Reynolds stress tensor at the inflow plane, which is interpolated by
a dataset of previous DNS of supersonic boundary layer performed by the same group~\citep{pirodb}.
The computation is initialized by prescribing a mean fully developed turbulent compressible boundary layer
obtained by applying the van Driest transformation~\citep{smits_dussauge06} to an incompressible profile of the Musker
family~\citep{musker79}.

\begin{figure*}
 \begin{center}
  \includegraphics[width=3.0cm,angle=270,clip]{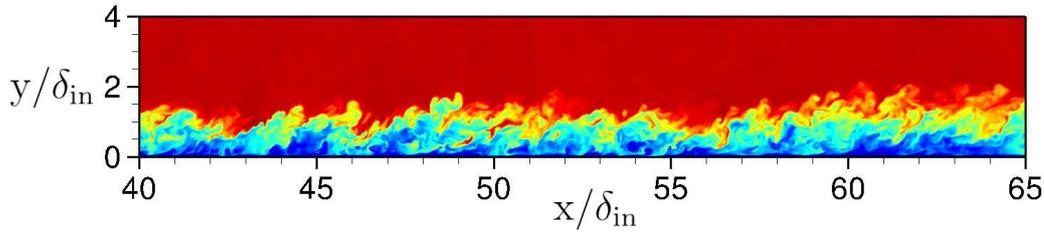} \vskip 0.5em
  \caption{\label{fig:plotxy} Instantaneous density field in a streamwise wall-normal plane.
           Contour levels are shown in the range $0.55 < \rho/\rho_{\infty}< 1.05$.}
 \end{center}
\end{figure*}

In figure~\ref{fig:plotxy} we show a snapshot of the instantaneous density field in a streamwise
wall-normal plane. The figure highlights the main features of the turbulent boundary layer and
its multi-scale structure, characterized by an extremely intermittent behavior in the outer layer,
with regions of relatively quiescent, high-speed irrotational fluid interspersed with slower,
large-scale rotational bulges. 
\begin{figure}
 \begin{center}
  \psfrag{x}[][][0.8]{$y^+$}
  \psfrag{y}[][][0.8]{$u_{VD}^+$}
  \includegraphics[width=3.75cm,angle=270,clip]{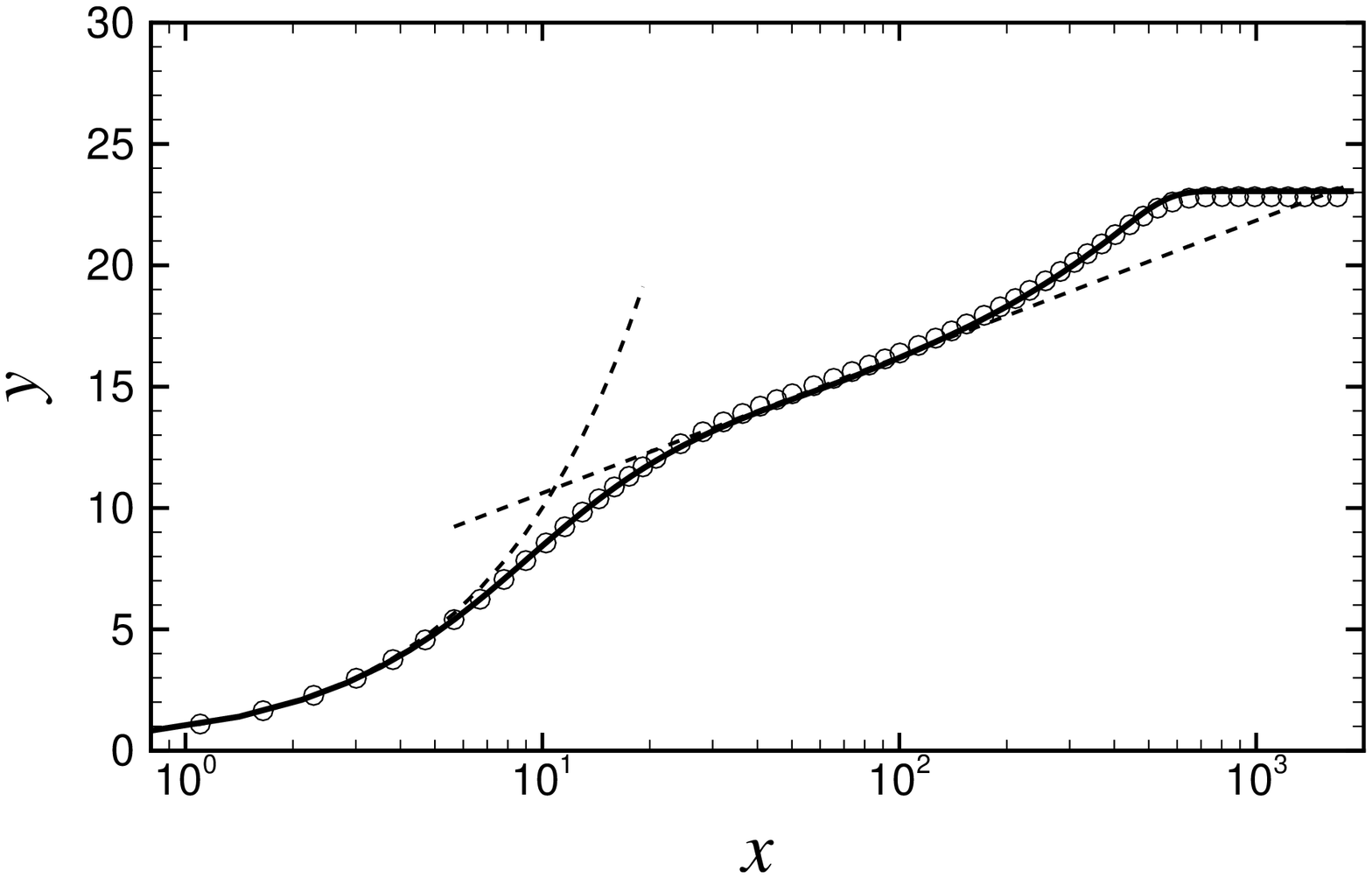} \hskip 1em
	 \psfrag{y}[][][0.8]{$\overline{\rho} \widetilde{u_i^{\prime \prime} u_j^{\prime \prime}} / \tau_w$}
  \includegraphics[width=3.75cm,angle=270,clip]{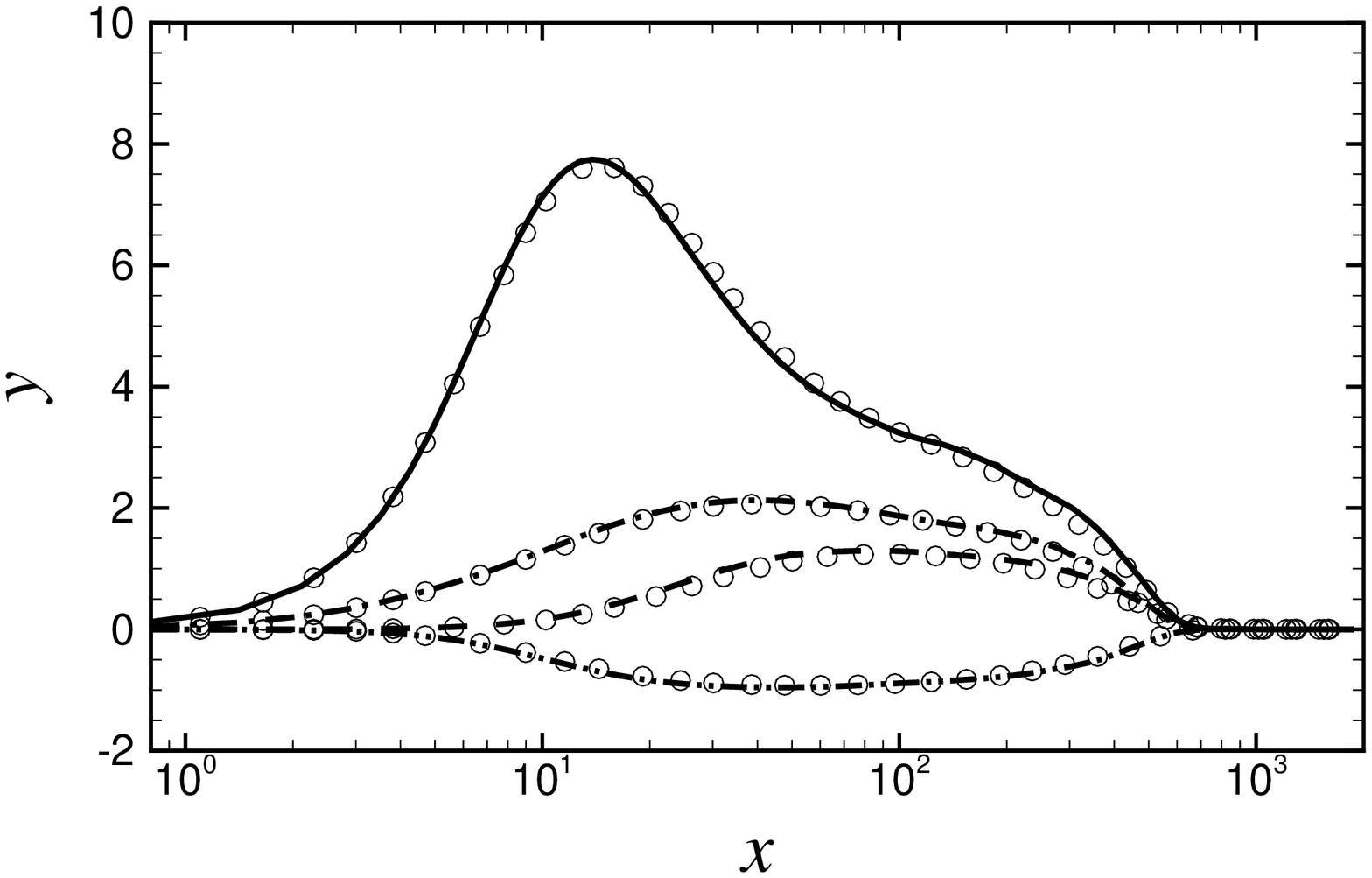} \vskip 0.5em
  \caption{\label{fig:blstat} Comparison of van-Driest transformed mean streamwise velocity (a) and fluctuating velocity statistics
 (b) scaled in wall units, with reference incompressible DNS~\citep{simens09,jimenez10} data at similar friction Reynolds number.
 Solid line, present DNS; symbols, reference data. The dashed lines in (a) denote
	 the linear $\overline{u}^+ = y^+$ and log-law $\overline{u}^+ = 5. + 2.44 \ln y^+$.}
 \end{center}
\end{figure}
The distributions of the van Driest transformed mean streamwise velocity profile
and velocity fluctuation intensities at a reference station ($x_{\textrm{ref}} = 90 \delta_{\textrm{in}}$)
are reported in figure~\ref{fig:blstat} in inner scaling. The DNS data are compared with
the incompressible boundary layer datasets~\citep{simens09,jimenez10} at similar friction Reynolds number
($Re \approx 580$). The figure shows near collapse of compressible and incompressible DNS data,
after density variations are accounted for.

\subsection{Shock-wave/turbulent boundary layer interaction}

\begin{figure*}
 \begin{center}
  \includegraphics[width=6.0cm,angle=270,clip]{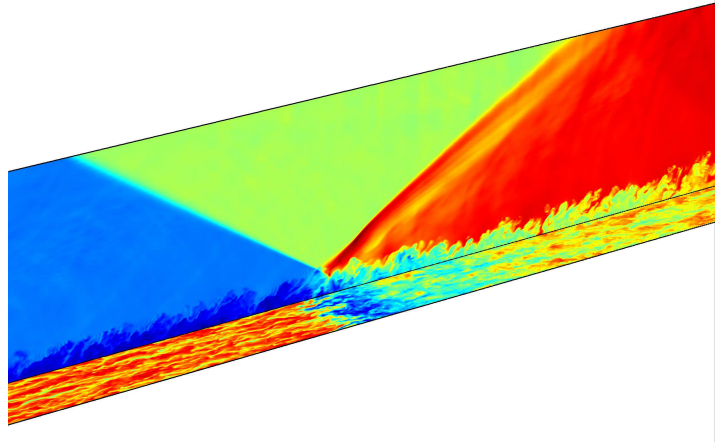} \vskip 0.5em
  \caption{\label{fig:3d_SBLI}Visualization of main SBLI features. Contours of the instantaneous density field in a streamwise wall-normal plane,
	   superposed with contours of streamwise velocity in a wall-parallel plane at $y^+ = 30$.}
 \end{center}
\end{figure*}
\begin{figure}
 \begin{center}
  \psfrag{x}[t][][0.8]{$(x-x_{\textrm{imp}})/L$}
  \psfrag{y}[b][][0.8]{$(\overline{p}_w-p_{\infty})/(p_2-p_{\infty})$}
  \includegraphics[width=3.75cm,angle=270,clip]{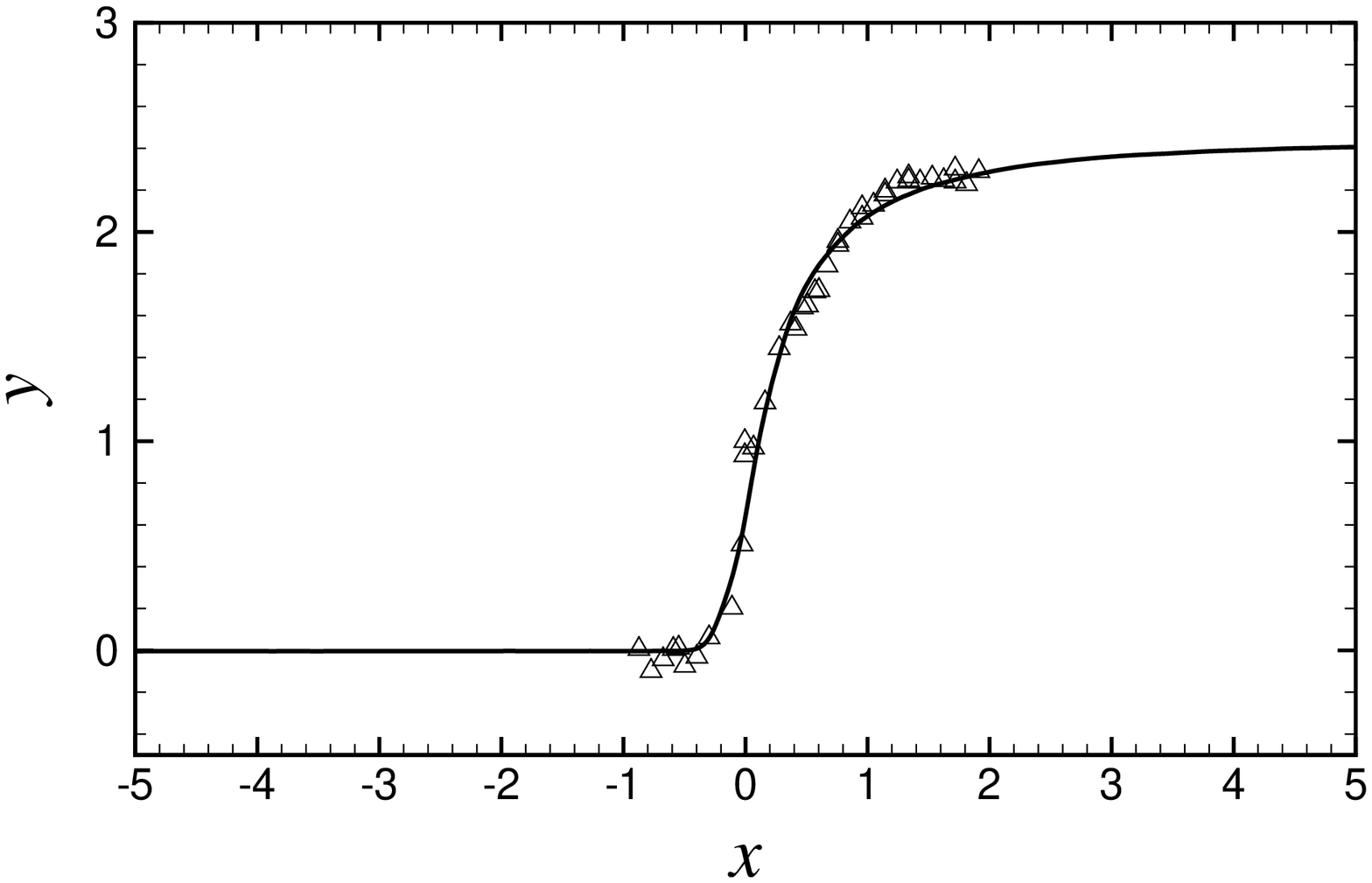} \hskip 1em
	 \psfrag{y}[b][][0.8]{$\sqrt{\widetilde{u'' u''}}/u_{\infty}$}
  \includegraphics[width=3.75cm,angle=270,clip]{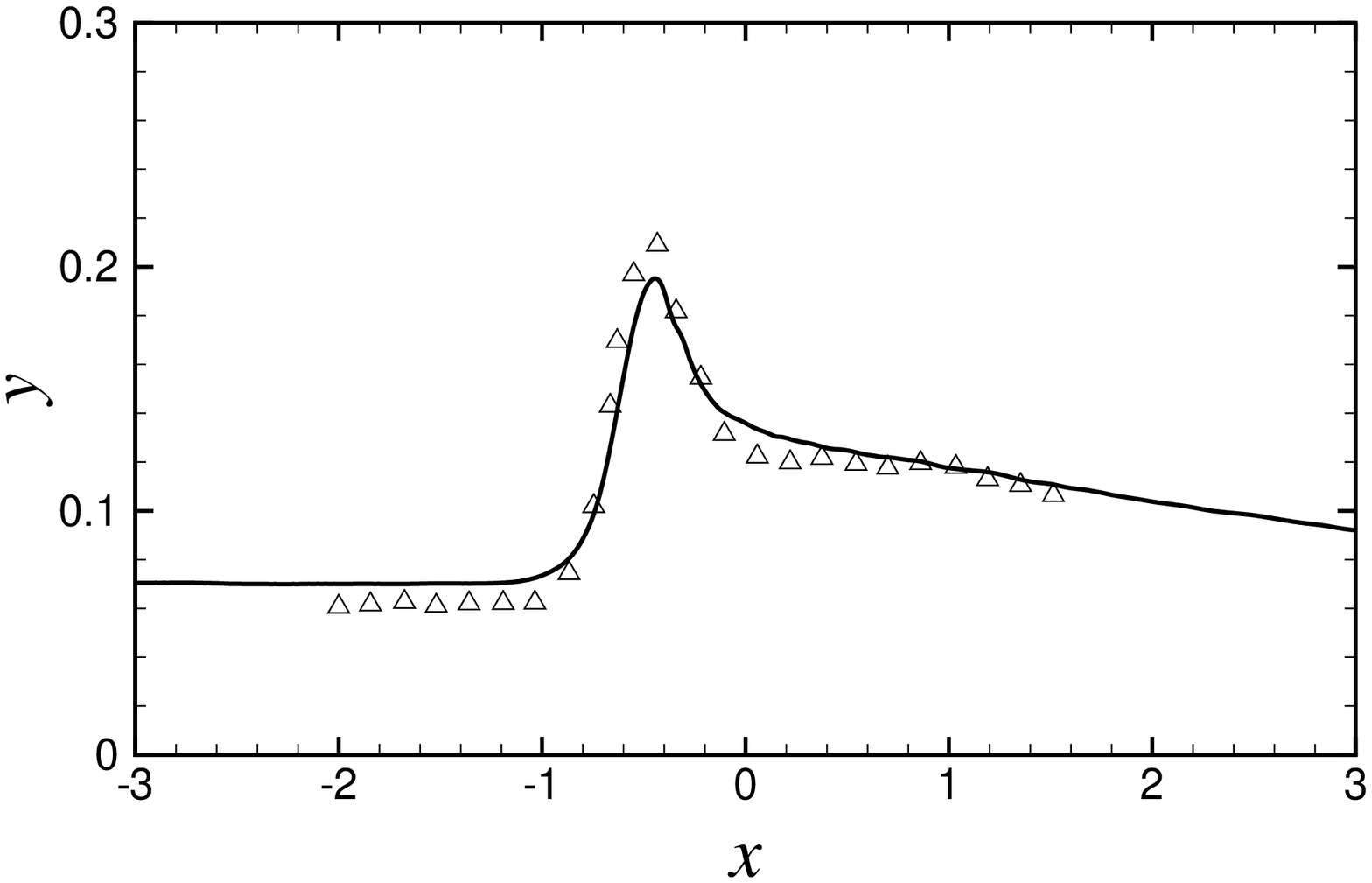} \vskip 0.5em
  \caption{Distribution of (a) mean wall pressure and (b) streamwise turbulent fluctuation intensity at $y = 0.1 L$ as a
	 function of the scaled interaction coordinate $(x-x_{\textrm{imp}})/L$.
	 Solid line, DNS data; open triangles, reference experiment~\citep{dupont06}.}
	 \label{fig:sblistat}
 \end{center}
\end{figure}
We present a third flow case to test the shock-capturing capabilities of STREAmS.
We carry out DNS of shock-wave/turbulent boundary layer interaction to replicate the flow conditions of 
reference experiments~\citep{dupont06}, characterized by a free-stream Mach number $M=2.28$ and
incidence angle of the shock generator $\phi = 8^\circ$. 

The simulation is performed in a computational domain of size
$L_x \times L_y \times L_z = [100\times 12 \times 6]\delta_{\textrm{in}}$
discretised using $N_x \times N_y \times N_z = [4096 \times 384 \times 288]$ grid points.
Here $\delta_{\textrm{in}}$ represents the thickness of the incoming boundary layer upstream of the interaction.
The specification of the boundary condition follows the setup adopted for the previous flow case,
except for the upper boundary of the computational
domain, where the shock is artificially generated by imposing the inviscid
oblique shock solution corresponding to the selected flow deflection.

The flow organization in the investigated SBLI is given by figure~\ref{fig:3d_SBLI}, where contours
of the density field are shown in a streamwise-wall-normal plane superposed
with contours of the streamwise velocity fluctuations in a wall-parallel plane.
The figure shows the complex structure of the interaction, characterized by the presence of an impinging and a reflected shock,
which cause thickening of the incoming boundary layer, and the formation of a small recirculation bubble.
The typical pattern of high- and low-speed streaks that characterizes the organization of the streamwise velocity
disappears across the interaction region, and reform towards the end of the computational domain, where
the boundary layer gradually relaxes to the equilibrium state.

A comparison of DNS data with the reference experiment is reported in Fig ~\ref{fig:sblistat}, where the distribution of the mean wall pressure
and of the streamwise fluctuation intensity is shown across the interaction zone, in terms of the scaled
interaction coordinate $(x-x_{imp})/L$, $L$ being the distance between the nominal impingement point of the incoming shock and the
apparent origin of the reflected shock. It turns out that the structure of the interaction zone is well captured by the simulation,
which predicts a wall pressure rise in excellent agreement with the available experimental data. Similarly, very good agreement
is observed for the root-mean-square of the streamwise fluctuation intensity, whose increase in the interaction region is associated
with the amplification of turbulence caused by the adverse pressure gradient imparted by the shock system.

\section{GPU Implementation and computational performance}

\subsection{Parallelization and GPU porting}


HPC is currently facing a major transition as the majority of 
systems in operation is still based on CPUs, but GPU-based systems are experiencing rapid growth.
For this reason in this phase it is very useful to have a code which can be used on different architectures 
without requiring further modifications. 
Tuning the code for different architectures typically involves considerable commitment,
including management effort in maintaining,
updating or modifying multiple versions of the same code. 
For this reason we design STREAmS to efficiently work on the most common HPC architectures operating today.
The code is written in the Fortran language -- mostly using Fortran 90 features --
which is widely used in HPC, and it is parallelized using the MPI paradigm.
Domain decomposition is carried out in two directions -- streamwise and spanwise -- in order to limit the
amount of data transferred for updating the ghost nodes,
considering that the communication times may become important when using a large number of tasks.

STREAmS has been developed to support the use of multi-GPUs architectures, while retaining the possibility to compile
and use the code on standard CPU based systems.
To achieve this goal, different programming approaches are possible. A first option is using directives, for instance
OpenACC~\cite{openacc} or OpenMP~\cite{openmp}, which allows to keep the CPU code completely unchanged.
A second approach relies on the use of specialized platforms for a specific hardware, which for NVIDIA GPUs
are CUDA \cite{cuda} and CUDA-Fortran \cite{cudafortran}. A third strategy is to use more portable
but more inconvenient or less popular tools,
such as OpenCL \cite{opencl} or HIP: C ++ Heterogeneous-Compute Interface for Portability \cite{hip}.

For these reasons in STREAmS we opt for CUDA-Fortran as this allows us
to achieve good parallel performance while limiting the changes to the initial CPU code.
In particular, the use of the \verb+cuf+ automatic kernels allows the large majority of the code to
remain unaltered, thus avoiding keeping different versions of the code.
The GPU-specific parts of the solver are marked by the \verb+#ifdef USE_CUDA+ preprocessing directive.
This strategy resembles the approach adopted by other popular codes in the field of
incompressible turbulence such as AFiD \citep{zhu_18} and CaNS \citep{costa_18}.

Another important part of the GPU porting is represented by the memory management between CPU and GPU. 
AFiD employs duplicated arrays residing on host and device, e.g. \verb+w+ and \verb+w_gpu+,
respectively. The device arrays are distinguished using the CUDA Fortran \textit{device}
attribute and are active only when CUDA compilation is enabled,
i.e. declared in modules inside preprocessing regions marked by \verb+USE_CUDA+ tokens.
When using device variables in the computing procedures, the variables are renamed inside
\verb+USE_CUDA+ regions using module aliasing so that the computations can always work with the
normal (host) names, i.e. \verb+use param, only: w => w_gpu+.
If the variables are passed, the declaration of the dummy arguments must also be
distinguished by adding \verb+attributes (device)+ inside \verb+USE_CUDA+ regions.
CaNS instead uses a more recent approach based on CUDA \textit{managed} memory.
The managed memory potentially allows to avoid completely the declaration of the CPU and GPU
versions of the same variable that can instead be used both in CPU and GPU code sections. 
However, the use of managed
memory requires particular care to optimize the underlying
transfers and to avoid undesired automatic transfers.
To achieve a good managed memory implementation, some information must be provided
to the CUDA platform, for example through the \textit{cudaMemAdvise}
and \textit{cudaMemPrefetchAsync} functions, which in our opinion reduce the readability of the code.
For this reason in STREAmS we followed a different approach, based on the following strategy. 
For each array, two versions are declared inside the Fortran module:
a baseline array \verb+w+ and the corresponding computing array \verb+w_gpu+. The latter
resides on the device, i.e. has the device attribute, only if the code is
compiled by activating CUDA.

\begin{verbatim}
real, allocatable, dimension(:,:,:;:) :: w, w_gpu
#ifdef USE_CUDA
attributes(device) :: w_gpu
#endif
\end{verbatim}

Moreover, \verb+w_gpu+ is explicitly allocated only if CUDA compilation is active.

\begin{verbatim}
#ifdef USE_CUDA
allocate(w_gpu(1:nx, 1:ny, 1:nz, 5))
#endif
\end{verbatim}

The baseline array \verb+w+ is used during the code initialization and finalization stages while \verb+w_gpu+ is used during the time marching section. 
To this aim, before starting the time evolution it is necessary to ensure that \verb+w_gpu+ contains the same data as \verb+w+. 
If CUDA is active, this is achieved by making a CPU-to-GPU copy managed transparently by CUDA-Fortran. 
If CUDA is not active, Fortran's \verb+move_alloc+ procedure is used, which allows to move the allocation from \verb+w+ to \verb+w_gpu+, both on CPU.

\begin{verbatim}
#ifdef USE_CUDA
w_gpu = w
#else
call move_alloc(w, w_gpu)
#endif
\end{verbatim}

A similar procedure is applied for the reverse transfer from GPU to CPU. 
In conclusion, with this memory management
the changes to the original solver are limited to variable declaration and allocation
and data transfer between CPU and GPU at the beginning and at the end of the computation,
while all the other parts of the code remain mostly unchanged.
Specification of the CUDA kernels is done by using
automatic cuf syntax,
\begin{verbatim}
!$cuf kernel do(3) <<<*,*>>> 
 do k=1,nz
  do j=1,ny
   do i=1,nx
    do m=1,nv
     w_gpu(m,i,j,k) = w_gpu(m,i,j,k)+fln_gpu(m,i,j,k)
    enddo
   enddo
  enddo
 enddo
!@cuf iercuda=cudaDeviceSynchronize()
\end{verbatim}
therefore only minor changes to the loops are needed. 
In particular, interdependent loops have been avoided as well as small loops, which have been
converted into scalar variables.

\begin{figure*}
 \begin{center}
  \includegraphics[width=0.9\textwidth,clip]{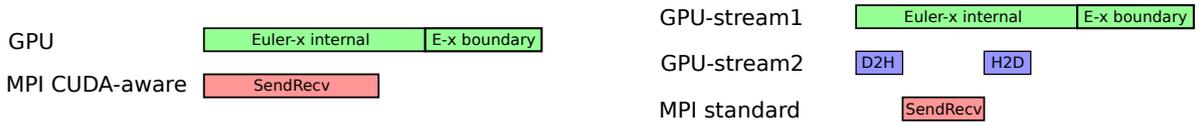} \vskip 0.5em
	 \caption{\label{fig:mpioverlap}Sketch of asynchronous time-lines for the evaluation of convective fluxes: 
	 internal fluxes evaluation (\textit{Euler-x internal}), boundary fluxes evaluation (\textit{E-x boundary}), 
	 host-to-device transfers (\textit{H2D}), device-to-host transfers (\textit{D2H}), MPI communications (\textit{SendRecv}).
	 CUDA-Aware MPI based time-line is shown on the left while standard MPI time-line is provided on the right. }
 \end{center}
\end{figure*}

Large computational domains require the use of multiple GPUs, in which each MPI process typically manages one graphic card.
Communication between multiple GPUs can be carried out in two main fashions. The first option relies on manual
copy between host and GPU to guarantee that the MPI communications always occur between variables residing on the host.
The second option relies on the so-called CUDA-Aware MPI implementations which allow the user to call MPI 
application programming interface (API)
passing device-resident variables. STREAmS has been parallelized to support both data communication patterns, selectable according
to compilation options. This allows to correctly run in environments where CUDA-Aware implementations are not available. 
MPI communications between multiple GPUs can negatively affect the parallel performance of the solver,
depending on the network speed between the computing nodes.
To improve the scalability performance, the GPU implementation
of STREAmS optionally supports asynchronous patterns
in which the GPU computation is overlapped with the swapping procedure necessary to exchange information across adjacent blocks.
This is done by exploiting the built-in asynchrony of the CUDA kernels and the capabilities of the CUDA \textit{streams}.
In this regard, two slightly different strategies were implemented depending on the
availability of the CUDA-Aware MPI. As an example, in Figure~\ref{fig:mpioverlap} a sketch of the time-lines corresponding
to the evaluation of the streamwise convective fluxes are represented. Basically, the evaluation for internal points 
can be performed before receiving the ghost nodes and for this reason can be overlapped with MPI communications. 
Following this idea, the CUDA-Aware MPI implementation (left) is straightforward while the standard MPI implementation 
(right) requires asynchronous CPU-GPU transfers using \textit{cudaMemcpyAsync} in a specific CUDA stream.
After receiving the data on ghost nodes, the boundary values can be computed.

\subsection{Performance results}

\begin{figure}
    \begin{center}
        \includegraphics[width=0.45\textwidth]{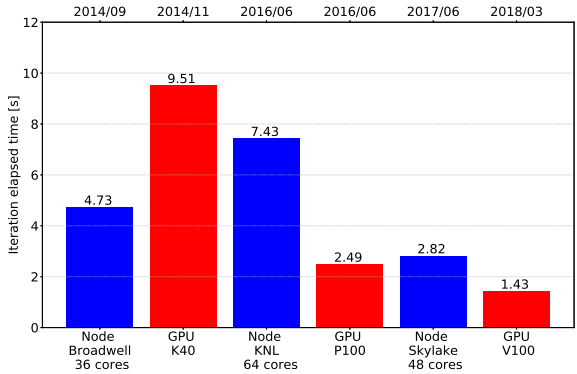}
	\caption{\label{perfsingle} Comparison of STREAmS performance -- elapsed time per iteration vs  -- using different 
	    HPC architectures: elapsed time per iteration versus the used processing unit is provided. 
	    For CPU-based runs a single computing node is employed using the MPI parallelization. 
	    For GPU-based runs a single GPU is used. The grid size is 1008$\times$251$\times$234 (1008$\times$251$\times$236 for 
	    KNL run). Improvement of processing unit power over years is underlined by the release dates in the upper horizontal axis.}
    \end{center}
\end{figure}

In this section we discuss the parallel performance
of STREAmS, reporting both weak and strong scaling of the code.
We have carried out test runs at CINECA on the HPC cluster DAVIDE,
an energy-aware Petaflops Class High Performance
Cluster based on Power 8 Architecture and coupled with NVIDIA Tesla
Pascal GPUs P100 with NVLink.
We first report the STREAmS performance on a single GPU card, 
comparing the elapsed time per time step to a single CPU node, for different HPC
architectures, figure~\ref{perfsingle}. For the CPU version of the code, we use the Intel compiler.
For this test we use a computational mesh with $1008\times251\times234$, corresponding
to about 10GB of memory allocation on a GPU. 
On the top axis figure~\ref{perfsingle} we report the release dates for each processing unit,
highlighting that GPUs present a more significant improvement over the years with respect to CPUs.

\begin{figure}
    \begin{center}
        \includegraphics[width=0.45\textwidth]{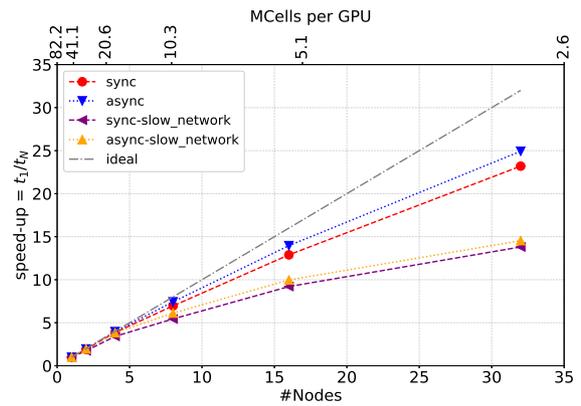}
	\caption{\label{perfstrong} Strong scaling of STREAmS using GPU P100 based cluster D.A.V.I.D.E.:
	    the speed-up of execution compared to the reference case is plotted against the number
	    of used nodes. The grid size is 2560$\times$251$\times$512. The reference case employs one computational
	    node equipped with 4 GPUs. On the upper horizontal axis the millions of cells residing on
	    each GPU is provided to understand the link between the GPU filling and the performances.
	    Four types of MPI parallelization are considered: synchronous, asynchronous, 
            synchronous with artificially slow network and asynchronous with artificially slow network.}
    \end{center}
\end{figure}

As for the solver performance on multiple nodes/GPUs
we note that communications between GPUs could result in lower performance
compared to CPUs, due to the additional data transfer between CPU and GPU.
On the other hand, this additional computational cost is typically mitigated
due to the use of fewer MPI processes when running on GPUs with respect to CPUs. 
Figure~\ref{perfstrong}
shows the speed-up of the synchronous and asynchronous versions of the code
keeping constant the number of grid points $2560\times251\times512$ and using 4 GPUs (one computing node) as reference.
Two additional speed-up curves are also reported, corresponding to artificially reduced
performance of the interconnection. Low-performance interconnection was emulated by
intentionally reiterating the MPI communications seven times. CUDA-Aware MPI was used.

Strong scalability is very good up to 4 computing nodes (16 GPUs) and good up
to 16 nodes. After that, the speed-up still increases but efficiency degradation
is significant. 
We find that for both network conditions 
the asynchronous version of the code is slightly faster than the regular one.
On the top axis we report the millions of cells processed
by each GPU which shows that 10 millions points per GPU are a reasonable
threshold to guarantee optimal efficiency.

\begin{figure}
    \begin{center}
        \includegraphics[width=0.45\textwidth]{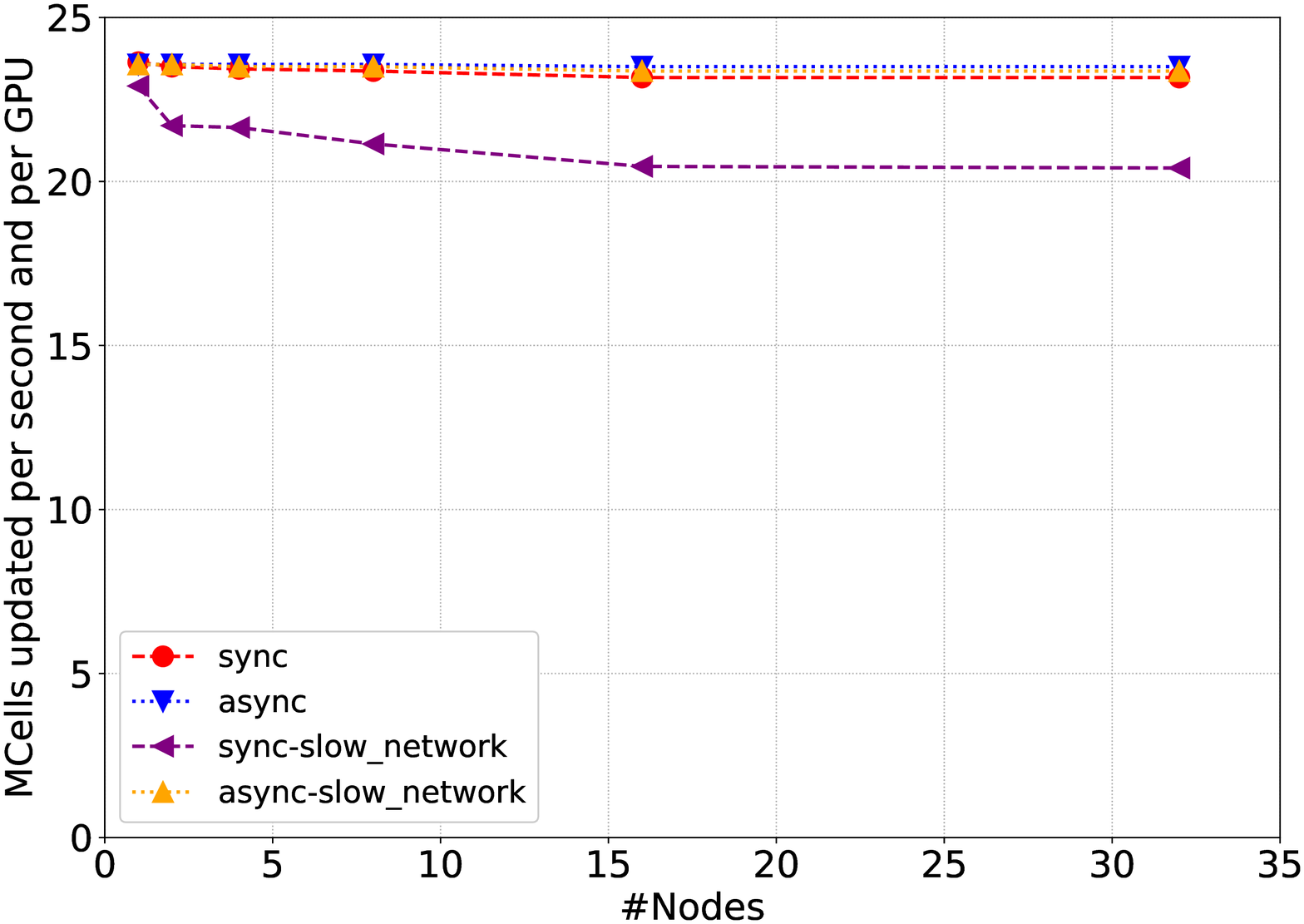}
	\caption{\label{perfweak} Weak scaling of STREAmS using the GPU P100 based cluster D.A.V.I.D.E.: the millions of cells
	    processed by each GPU per second are plotted against the number of computing nodes.
            Four types of MPI parallelizations are considered: synchronous, asynchronous, synchronous with artificially slow network and asynchronous with artificially slow network.}
    \end{center}
\end{figure}

In Figure~\ref{perfweak} we provide a measure of the weak-scaling performance, i.e. by
keeping constant the number of cells processed by each GPU. 
The reference case has again a mesh with $2560\times251\times512$ points, but the grid is scaled
as the number of nodes increases. We report the number of cells
updated per second and per GPU, which should be constant when 
increasing the total number of GPUs.
The performance of the synchronous and asynchronous version of 
the solver for the case with the real (fast)
network are both excellent. In the case of artificially slow network, the synchronous version
shows performance degradation, whereas the asynchronous code 
completely hides the communication times. It may be concluded that using
the asynchronous version of the code is always recommended, although the
performance improvement is not always significant.

Given the availability of recently developed GPU solver for the simulation of canonical incompressible turbulent flows, i.e.
AFiD and CaNS, it is interesting to attempt a comparison with the STREAmS performance.
The comparison is physically significant in the low-Mach-number limit (say $M_{\infty}<0.2$),
at which the results of incompressible and compressible solvers are basically identical~\citep{modesti_18a}.
Obviously, compressible solvers require the evaluation of a much more complex
right-hand side and use of smaller time step owing to the acoustic time step restriction.
On the other hand, incompressible solvers requires the solution of a Poisson
equation at each time step, which is the most computation-intensive part
of the solver, implying extensive use of all-to-all MPI communications.
AFiD solves the incompressible Navier-Stokes equations with the Boussinesq
approximation for temperature with implicit treatment of the diffusive
terms. CaNS solves the incompressible equations
without the scalar equation, and can be compiled with explicit or implicit
treatment of the diffusive terms. The measured clock times per iteration 
of STREAmS, AFiD and CaNS are compared in Table~\ref{perftable}, in a weak scaling
test with base computational mesh of 8.39 million of points, and with up to 2048 GPUs. 
The benchmarks were performed using the CSCS Piz-Daint cluster featuring one P100 GPU per node.
Asynchronous standard MPI compilation of STREAmS was employed.

\begin{table*}
	\centering
\begin{tabular}{@{}c|c|c|c|c|c|c|c@{}}
	\# GPUs  &     1     &    2    &    8     &    32    &    128    &    512    &    2048    \\ \midrule
	STREAmS  &     3.65  &    3.65 &    3.65  &    3.65  &    3.65   &    3.65   &    3.65    \\ 
	AFiD     &    0.84   &   1.59  & 1.76     & 2.17     & 3.19      & 5.5       &     6      \\
	CaNS     &     0.36  & 1.67    & 2.1      & 1.37     & 1.8       & 2.34      &     -
\end{tabular}
	\caption{\label{perftable} Performances of STREAmS compared to popular CFD CUDA-enabled codes AFiD and CaNS.
	Seven computational grids are provided in the weak scaling spirit starting from 8.39 million grid points case. 
	Correspondingly, the number of employed GPUs ranges from 1 to 2048. For CaNS, explicit compilation for diffusive 
	terms is considered.}
\end{table*}


Although actual code comparison 
should also consider the accuracy of the numerical methods
needed to achieve equivalent results, it is possible to
attempt an interpretation of the trends.
First it is observed that, when using a small number of GPUs, the time required
by STREAmS for single iteration is larger that for AFiD and for the explicit version of CaNS.
On the other hand, since STREAmS shows nearly ideal weak scalability,
the performance gap with the incompressible solvers is reduced as the number of GPUs
increases, eventually reversing for the AFiD implicit diffusion code. 
Albeit in a limited and partial context, and depending on the Mach number, 
this trend suggests that compressible solvers can potentially become competitive
with classical incompressible solvers in massively parallel calculations 
on HPC platforms with thousands of GPUs.

\section{Conclusions}

We heve presented a recent version of our in-house compressible flow solver STREAmS, that has been 
been ported to CUDA-Fortran and tailored to canonical wall-bounded flows, namely compressible channel,
supersonic boundary layer and SBLI.
STREAmS stems from two decades experience of our research group on DNS of compressible wall-bounded flows
and a baseline version of the solver, is released open-source under GPLv3 
license with the aim to provide the fluid dynamics community with a highly-parallel compressible flow solver.
The use of CUDA-Fortran with the use of \verb+cuf+ automatic kernels allowed us to largely minimize 
the changes to the original flow solver and to compile and run the code on different HPC architectures.
The tests carried out on the GPU cluster DAVIDE at CINECA show very good scalability performance,
proving that the solver can be used to carry out large scale direct numerical simulations. 
Interestingly STREAmS shows improved weak scalability performance compared to the state-of-the-art incompressible GPU
solvers. Although it is well known that compressible flow solvers typically show
better scalability than the incompressible ones, in our experience the difference in performance is less marked on CPUs.
This preliminary results show that for large scale simulations using thousands of GPUs
the use of compressible flow solvers operating at low Mach number could potentially become competitive
with incompressible solvers, despite the substantial overhead in terms of floating point operations, 
and the restriction on the acoustic time step limitation.

The availability of the GPU version of the solver will allow to take advantage of the contemporary
pre-Exascale systems and the next generation of Exascale supercomputers currently under development,
allowing to significantly extend the range of simulated Reynolds number up to
the genuine high-Reynolds number regime ($Re_{\tau} > 10^4$).
This opportunity will allow the flow community to provide definite answers to key issues, 
as the presence of a logarithmic range of variation of the streamwise velocity
variance with the wall distance, as predicted in the overlap layer by the
attached eddy hypothesis, and for which partial support comes from high-Reynolds-number experiments.

{\bf Acknowledgements}\\
M. Bernardini was supported by the Scientific Independence of Young Researchers program 2014 (Active Control of Shock-Wave/ Boundary-Layer
Interactions project, grant RBSI14TKWU), which is funded by the Ministero Istruzione Universit\`a e Ricerca. The authors are especially
grateful for the computational resources provided by the Cineca Italian Computing Center.

\printcredits
\bibliographystyle{elsarticle-num}
\bibliography{paper}   

\end{document}